\documentclass[journal=nalefd,manuscript=article]{achemso}

\usepackage[version=3]{mhchem} % Formula subscripts using \ce{}
\graphicspath{{Figures/}} %add
\usepackage{xcolor} %add
\author{Georgy V. Pushkarev}
\author{Vladimir G. Mazurenko}
\author{Vladimir V. Mazurenko}
\author{Danil W. Boukhvalov}
\email{danil@njfu.edu.cn}

\affiliation[Un1]
{Ural Federal University, Ekaterinburg, 620002 Russia.}
\alsoaffiliation[Un2]
{College of Science, Institute of Materials Physics and Chemistry, Nanjing Forestry University, 210037, Nanjing, PR China}

\title[An \textsf{achemso} demo]
  {On the Nature of Interlayer Bonds in Two-dimensional Materials}

\abbreviations{2D,DFT,hBN,WF}

\begin{document}
\begin{abstract}
The role of interlayer bonds in the two-dimensional (2D) materials "beyond graphene" and so-called van der Waals heterostructures is vital, and understanding the nature of these bonds in terms of strength and type is essential due to a wide range of their prospective technological applications. However, this issue has not yet been properly addressed in the previous investigations devoted to 2D materials. In our work, by using first-principles calculations we perform a systematic study of the interlayer bonds and charge redistribution of several representative 2D materials that are traditionally referred as van der Waals systems. Our results demonstrate that one can distinguish three main types of inter-layer couplings in the considered 2D structures: one atom thick membranes bonded by London dispersion forces (graphene, hBN), systems with leading electrostatic interaction between layers (diselenides, InSe and bilayer silica) and materials with so-called dative or coordination chemical bonds between layers (ditelurides). We also propose a protocol for recognising the leading type of interlayer bonds in a system that includes comparison of interlayer distances, binding energies and redistribution of the charge densities in interlayer space. Such an approach is computationally cheap and can be used to further prediction of chemical and physical properties such as charge density waves (CDW), work function and chemical stability at ambient conditions.
\end{abstract}

%%%MAIN TEXT%%%%
\section{Introduction}
Two-dimensional materials beyond graphene attract much attention
in last decade. \cite{I1,I2,I3,I4,I5,I6,I10,I11} These structures are usually constructed from several layers connected with non-covalent interlayer bonds. So-called van der Waals heterostructures constructed from various monolayer \cite{I7,I8,I12,I13} are also considered as key materials with multiple prospective applications in electronics \cite{Briggs2019,Kang2020,Kim2021}, energy storage \cite{Ying2019,Yanfeng2017,Shi2017}, electro-\cite{Fu2017,Tang2019,Su2019} and photo–catalysis \cite{Luo2016}, sensing \cite{Anichini2018,Zhou2019,Munteanu2021}, magnetism \cite{Wang2022} etc.  

Results obtained in experimental studies  \cite{Hossain2012, Liu2008} and theoretical simulations \cite{Boukhvalov2010} demonstrate a significant influence of the number and quality of subsurface layers on chemical activity of surface of few–layer graphene systems and  other 2D materials \cite{I9,Edla2020,D'Olimpio2022,Ahmad2019}. Another important example that evidences a significant role of the interlayer interactions is discrepancy between chemical stability of InSe monolayer and chemical instability of the surface of the same compound in the bulk form \cite{D'Olimpio2020,Politano2016}. Thus, unveiling the nature of  non-covalent bonds between the neighbouring layers is essential for understanding of physical and chemical properties of these structures and further intelligent design of novel layered materials. Surprisingly, while 2D materials were systematized by chemical composition (for example monochalcogenides, dichalcogenides, MXenes) or by chemical composition in numerous reviews, differences in strength and types of interlayer bonds were not discussed and used for description of 2D materials.\cite{Mas-Balleste2011,Ankur2015,Li2017,Glavin2020,Lam2022}

In this work we report a systematic study of the bonds length and strength as well as charge distribution in bulk and bilayer structures of several representative 2D systems such as graphene, hexagonal boron nitride (hBN), MoSe$_2$, VSe$_2$, and ditelureides (NiTe$_2$, PdTe$_2$, PtTe$_2$) by means of the first-principles calculations. Graphene and hBN have been chosen as classical examples of one-atom-thick crystals,  MoSe$_2$ being the most studied material from dichalcogenides family, VSe$_2$ that represents another popular diselenides with \textit{3d} metal center and nonzero magnetic moment \cite{Bonilla2018,ma2012,Boukhvalov2020} and NiTe$_2$, PdTe$_2$, PtTe$_2$ as three ditelurides with the same morphology of the layer and metal center from the same group of the periodic table. To make list of the considered systems more diverse, we also explore two-layer materials in which each layer is constructed from two sheets connected via covalent bonds. The first material is indium selenide (InSe) \cite{Man1976}. Unique values of direct and indirect band gaps in the electronic spectrum make this material suitable for solar energy applications \cite{Segura1983}. Dependence of the electronic structure of  few-layer InSe crystals on the sample thickness revealed in previous experimental \cite{Lei2014,Brotons-Gisbert2016} and theoretical\cite{Brotons-Gisbert2016,Zolyomi2014} works also suggests these systems to be promising for optoelectronics. The second material is silica or silicon dioxide bilayer with stable 2D structure modifications. The lowest energy was reported for the $\alpha$-2D-silica\cite{Gao2017,Huang2012,Loffler2015,Altman2013,Hutchings2022}. The band gap value of about 7 eV \cite{Gao2017} that is the largest one among 2D insulating materials makes this material a perspective candidate for future applications in nanoelectronics and for exploring fundamental physical phenomena \cite{BUCHNER2017}. Based on the simulated properties we perform classification of the considered systems with respect to the type of interlayer chemical bonds. Importantly, in this study we do not consider a number of actual two-dimensional materials such as  silicene\cite{Silicene_ZHAO2016,Silicene_Molle2018}, germanene \cite{Germanene_Davila_2014,Germanene_Suzuki2021} and borophene \cite{Borophene_Mannix2015,Borophene_Wenbin2018}, since their stabilization in experimental conditions is possible with metal substrates, which requires taking into account metallic bonds and therefore is out of scope of our investigation.

%%%%%%%%%%%%%%%%%%%%%%%%%%%%%%%%%%%%%%%%%%%%%%%%%%%%%%%%%%%%%%%%%%%%%%%%%%%%%%%%%%%%%%%%%%%%%%
%%%%%%%%%%%%%%%%%%%%%%%%%%%%%%%%%%%%%%%%%%%%%%%%%%%%%%%%%%%%%%%%%%%%%%%%%%%%%%%%%%%%%%%%%%%%%%
%%%%%%%%%%%%%%%%%%%%%%%%%%%%%%%%%%%%%%%%%%%%%%%%%%%%%%%%%%%%%%%%%%%%%%%%%%%%%%%%%%%%%%%%%%%%%%
\section{Computational Methods}
The simulations of atomic and electronic structure of all considered materials were carried out within Density Functional Theory (DFT) framework using the Perdew-Burke-Ernzerhof (PBE) exchange-correlation functional \cite{PBE} as implemented in the Vienna ab-initio simulation package (VASP) \cite{VASP1,VASP2} with a plane-wave basis set. We also included van der Waals interaction using the method of Grimme (DFT-D3) \cite{VdW}. The calculation parameters were chosen as follows. The energy cutoff equals to 500 eV and the energy convergence criteria is $10^{-8}$ eV. For the Brillouin zone integration a $20\times20\times1$ gamma centered grid was used for layered structures and $8\times8\times8$ for bulk structures. In the case of mono- and bilayers vacuum space more than 10 \AA \ between the layers related by periodic boundary conditions was introduced. 

We performed relaxation of atomic positions, calculations of electronic structure and charge densities distributions of all the considered systems. The structure data used in our calculations (parameters of the unit cells and atomic positions) are reported in the Supplementary Information. These first-principles results were used for estimation of the binding energy per interatomic bond that is given by
\begin{equation}
E_{\rm bind} = \frac{E_{\rm mono}-E_{\rm bulk}/n}{m},
\label{eq1}
\end{equation}
where $E_{\rm mono}$ and $E_{\rm bulk}$ are the total energies of monolayer and bulk supercells, correspondingly. $n$ denotes the number of layers in the unit cell used to simulate the bulk phase. This definition of the binding energy was also used in previous numerical simulations for characterizing individual 2D systems \cite{Thrower2013,Izgorodina2009,Habibi2021,Li2018}, but not for their comparison. According to Eq.\ref{eq1} the positive value of $E_{\rm bind}$ means that the formation of a bulk structure is more energy efficient than the monolayer one. As we will show below, this is the case for all the systems in question, which indicates that an additional energy is needed to stabilize a monolayer structure. The value of $m$ is the total number of the inter-layer non-covalent bonds per formula unit in the bulk system. In the case of the noble gases that are used as the examples of genuine van der Waals systems, we take $m=1$. Fig. S1 in the Supplementary Information gives a graphical representation of these bonds for all the systems in question. For layered systems with hexagonal and trigonal  lattices $m$ = 6. In the case of MoSe$_2$ and InSe, each Se atom belonging to the particular layer has three nearest neighbors from the next layer above or below, which makes the total number of inter-layer non-covalent bonds per the formula unit equal to six. For silica the number of such inter-layer bonds per the formula unit can be estimated to be 6. The main reason to use such a renormalization of the calculated values of $E_{\rm bind}$ with $m$ is the possibility to classify the systems in question with respect to the strength of the binding energy. It is also important for comparison with results obtained for atomic and molecular clusters that are referred in the literature as classical van der Waals systems and discussed in the next section. 

Another quantity of our interest that can be calculated from first-principles and allows us to classify two-dimensional materials is the work function, $\Psi_{\rm work}$. It corresponds to the energy needed to move electron from the surface of material to vacuum. $\Psi_{\rm work}$ can be calculated by using the following expression:
\begin{equation}
\Psi_{\rm work} = E_{\rm vac}-E_{\rm Fermi},
\label{eq2}
\end{equation}
where  $E_{\rm vac}$ is the energy of vacuum determined from the local potential function with in-plane degrees of freedom integrated out and taken at middle position between the slab and its replica in the neighbouring supercell. In our study the latter is varied from 5 \AA \ to 7 \AA \ depending on the considered system. In turn, $E_{\rm Fermi}$ is the Fermi energy calculated from self-consistent DFT calculation (see for details Ref. \citenum{Singh-Miller2009}). 

To quantitatively estimate the charge redistribution corresponding to formation of the bulk structure from the individual monolayers we have calculated the charge density difference  defined as
\begin{eqnarray}
\Delta \rho = \rho_{\rm bulk} - \sum_{i}^{n} \rho_i.
\label{density}
\end{eqnarray}
Here $\rho_{\rm bulk}$ corresponds to the electron density of the supercell having periodic boundary conditions, containing several layers and representing the bulk crystal in the first-principles calculations. In turn, $\rho_{i}$ is the electron density of the $i^{t h}$ monolayer isolated from the rest of the supercell. We would like to stress that in DFT simulations the calculated charge density function doesn't reach the exact zero value even away from atomic centers. For the sake of visualization it is a standard practice to neglect such a background charge density. In Vesta software \cite{VESTA} used in our work this parameter named "Isosurface Level". In our study we use complete charge density functions to calculate $\Delta \rho$ and explicitly report the value of the cut-off charge density, $\rm BG$ used to visualize $\Delta \rho$.

%%%%%%%%%%%%%%%%%%%%%%%%%%%%%%%%%%%%%%%%%%%%%%%%%%%%%%%%%%%%%%%%%%%%%%%%%%%%%%%%%%%%%%%%%%%%%%
%%%%%%%%%%%%%%%%%%%%%%%%%%%%%%%%%%%%%%%%%%%%%%%%%%%%%%%%%%%%%%%%%%%%%%%%%%%%%%%%%%%%%%%%%%%%%%
%%%%%%%%%%%%%%%%%%%%%%%%%%%%%%%%%%%%%%%%%%%%%%%%%%%%%%%%%%%%%%%%%%%%%%%%%%%%%%%%%%%%%%%%%%%%%%
\section{Results}
\subsection{Benchmark of van der Waals system}
To obtain proper theoretical descriptors for the materials characterized by van der Waals interaction, we performed calculations for the systems considered as textbook examples of this type of coupling: noble gases and benzene molecule. For them we have analyzed the interactions of two atoms (for Xe and Ne) and two benzene molecules attached normally as it was discussed in the literature (see Ref. \citenum{Maranzana2013}). For all the considered systems we optimized atomic positions and then visualized the change of the charge density after formation of non-covalent bond. 
Results of the calculations are shown in Fig. \ref{fig1}a-c. The obtained values of the charge density difference do not exceed 10$^{-3} e^{-}$/\AA$^3$. In turn,  the binding energies are equal to 1.41, 5.00 and 13.04  meV/mol for He, Ne, benzene, respectively. The equilibrium distance between two benzene molecules is 2.40 \AA. These values are rather close to the numbers reported in earlier simulations \cite{vonLilienfeld2004}. Importantly, in the case of the noble gas atoms the redistribution of the charge density is considerably smaller (see green ''cloud'' in Fig. \ref{fig1}a,b) than that resulted from benzene–benzene bonds formation (see Fig. \ref{fig1}c). Such a difference by one order of magnitude can be explained by the contribution of  $\pi$-orbitals of $sp2$-hybridized carbon atoms to the formation of non-covalent bonds.
The obtained values of the binding energies and charge density differences will be used as a benchmark in searching for true van der Waals bonds in the analysis below.

%%%%%%%%%%%%%%%%%%%%%%%%%%%%%%%%%%%%%%%%%%%%%%%%%%%%%%%%%%%%%%%%%%%%%%%%%%%%%%%%%%%%%%%%%%%%%%
%%%%%%%%%%%%%%%%%%%%%%%%%%%%%%%%%%%%%%%%%%%%%%%%%%%%%%%%%%%%%%%%%%%%%%%%%%%%%%%%%%%%%%%%%%%%%%
%%%%%%%%%%%%%%%%%%%%%%%%%%%%%%%%%%%%%%%%%%%%%%%%%%%%%%%%%%%%%%%%%%%%%%%%%%%%%%%%%%%%%%%%%%%%%%

\subsection{Classification of 2D Systems}
At the first step of our investigation we calculated atomic structure and interlayer binding energies for the whole list of the studied systems. The calculated structural parameters such as lattice constants, inter-atomic and interlayer distances (Table \ref{tab1}) agree with those measured and/or calculated in the previous experimental and theoretical works, which provides the necessary basis for further analysis of the electronic structures. 

Results of our calculations demonstrate that the estimated values of the binding energies between adjacent layers vary significantly across the list of the considered materials nominated in the literature as van der Waals systems. By using the obtained $E_{\rm bind}$ one can distinguish three groups of materials. The first one contains graphene, hBN and a composite graphene-hBN systems whose values of the binding energies are about 15 meV per interatomic bond. The single-bond binding energies for graphene, hBN are rather close to the values of 1-13 meV calculated for the systems with London dispersion forces based bonds discussed in the previous section and characterized by the $\sim r^{-6}$ dependence of the interaction energy on the distance between the entities \cite{Israelachvili2011}. Contrary, the simulated diselenides (MoSe$_2$ and VSe$_2$ ) and InSe and $\alpha$-2D-silica are characterized by $E_{\rm bind}$ almost three times larger than that for the first group. Following the results reported in Refs.\citenum{Larson1984,Elangannan2011} we can conclude that by the order of magnitude of the calculated $E_{\rm bind}$ ($\sim$ 50 meV) this second group is close to systems with hydrogen bond. In general, this type of bonding is originated from  electrostatic (dipole–dipole) interactions between positively and negatively charged parts of interacting systems. In turn, in the case of the ditelurides (NiTe$_2$,PtTe$_2$, PdTe$_2$) the maximal values of $E_{\rm bind}$ are larger than 100 meV per interatomic bond and close to those estimated for the bonds in systems with coordination (also called dative) bonds\cite{Muller1994} between metal centres and ligands or molecules. This latter type of chemical bonds is related to overlap between fully occupied orbitals of one atom and unoccupied orbitals of other. Traditionally, such bonds are discussed as one to be intermediate between covalent and non-covalent coupling.\cite{Muller1994} 

The consideration of the investigated materials within three distinct groups defined with the values of the binding energy perfectly matches with a visible difference in interlayer distances: about 3.5 {\AA} for materials with van der Waals bonds, 3.2 {\AA}  for systems with electrostatic bonds, and about 2.5 {\AA}  for materials with coordination bonds. To reveal the sensitivity of the atomic structure to the formation of the interlayer non-covalent bond we compared the optimized lattice constants of monolayers, bilayers and bulk structures for all the discussed materials. Results of the calculations (Table \ref{tab1}) demonstrate that for the compounds in which layers weakly connected by the forces similar to London dispersion ones, monolayer to bilayer and monolayer to bulk transitions almost do not affect the values of in-plane lattice constant (the corresponding changes are of order of 0.001 {\AA}). For the systems with presumably electrostatic (hydrogen-like) bonds formation of layered structures is related to much larger changes in in-plane lattice constants that are about of 0.02 {\AA}. Stabilization of coordination-like bonds between layers also provides a visible increase of the in-plane lattice constant by about 0.1 {\AA}. Thus the proposed division of the studied materials in three distinct group with different types of interlayer bonds is also confirmed by the unambiguous signatures in their structural properties.

\begin{figure*}[t]
    \centering
    \includegraphics[width=0.99\columnwidth]{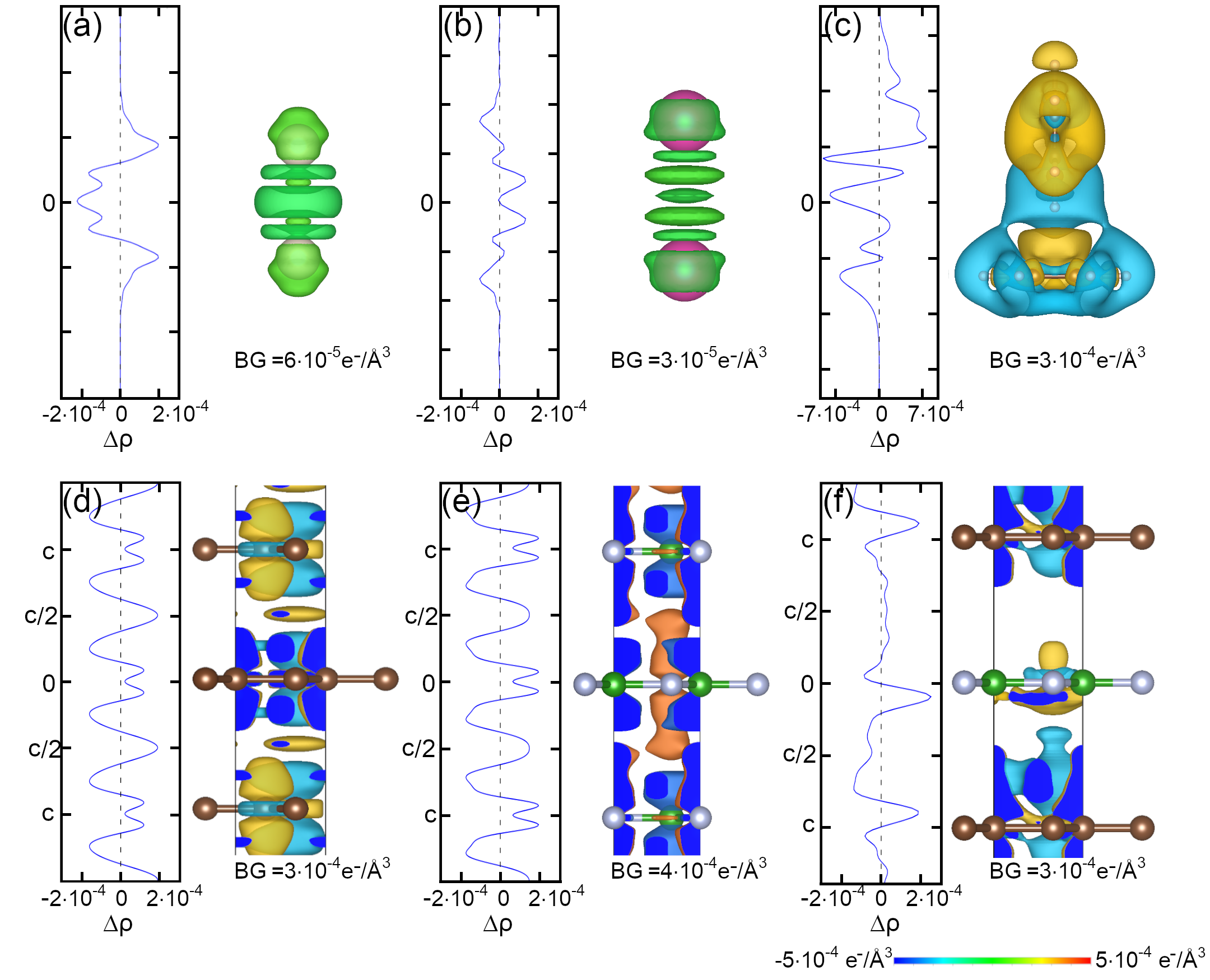}
    \caption{Numerical and visual representations of the  differences between charge densities before and after formation of noncovalent bonds between for two atoms of He (a), Ne (b), two benzene molecules (c), and bulk and monolayer of graphene (d), hexagonal boron nitride(e) and layers of h-BN alternating with layers of graphene (f). Numerical data plots demonstrate dependencies of the charge density differences, Eq.\ref{density} integrated over in-plane variables on the position in the vertical $z$ direction. All integrated density differences are presented in units of $e^{-}/$\AA. The background cut-off values for charge densities are different for each panel and differ from those presented in Fig. 2 and Fig. 3.
}
    \label{fig1}
\end{figure*}

\begin{figure*}[t]
    \centering
    \includegraphics[width=0.99\columnwidth]{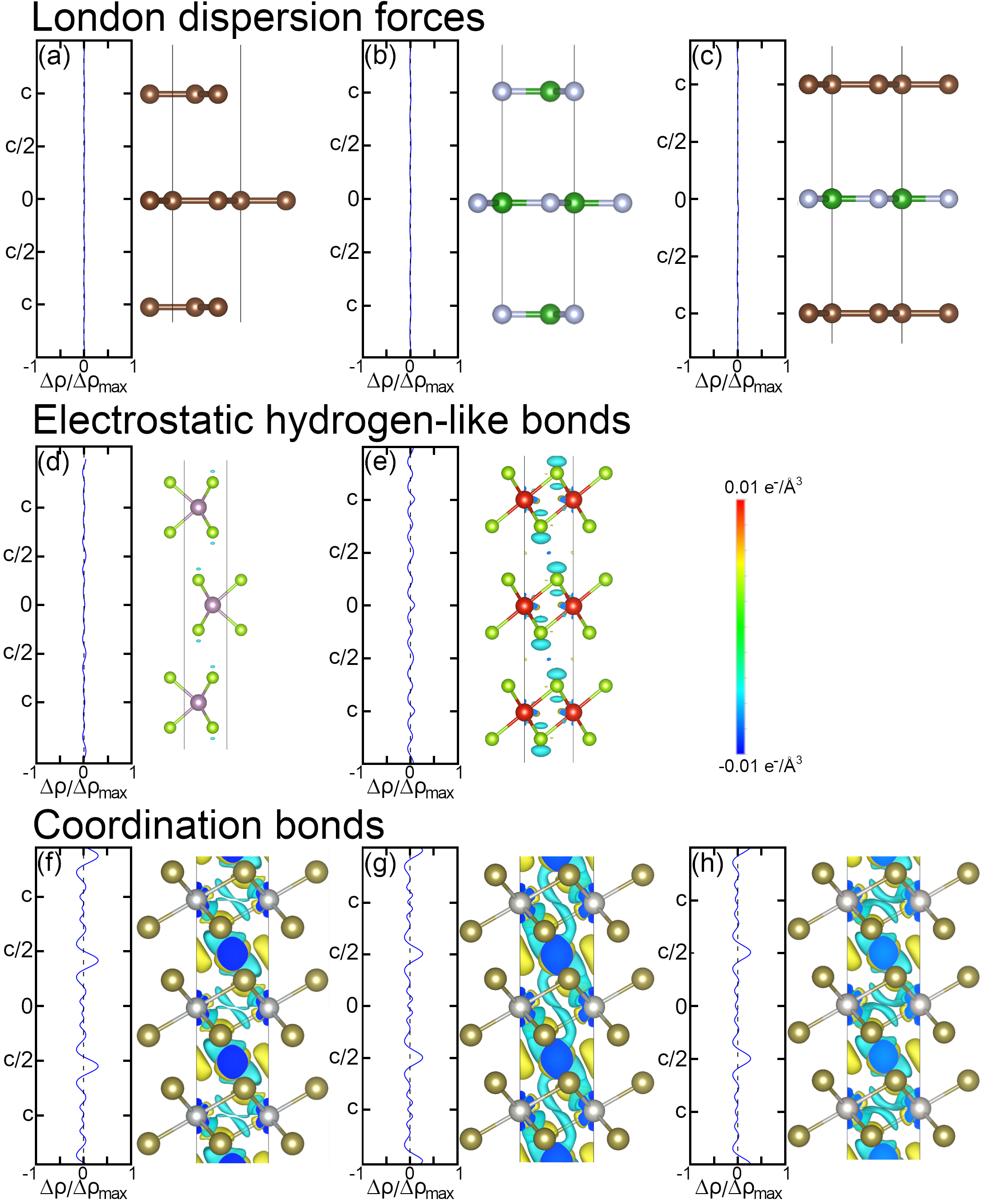}
    \caption{Numerical and visual representations of the  differences between charge densities of bulk and monolayer phases of graphene (a), hexagonal boron nitride (b), layer of h-BN alternating with layer of graphene(c), H-MoSe$_2$ (d), T-VSe$_2$ (e), NiTe$_2$ (f), PdTe$_2$ (g), PtTe$_2$ (h). Note that the background cut-off,${\rm BG}=5\times10^{-3} e^-/$\AA$^3$~ for charge densities changes presented numerically in graphs and visualized with Vesta package are the same for all panels of this figure and Fig. 3 but different from those presented in Fig. 1. The maximal value $\Delta \rho_{max}$ equals to 1.6 $\times$ 10$^{-2}$ $e^{-}/$\AA.
}
    \label{fig2}
\end{figure*}

\begin{figure*}[t]
    \centering
    \includegraphics[width=0.99\columnwidth]{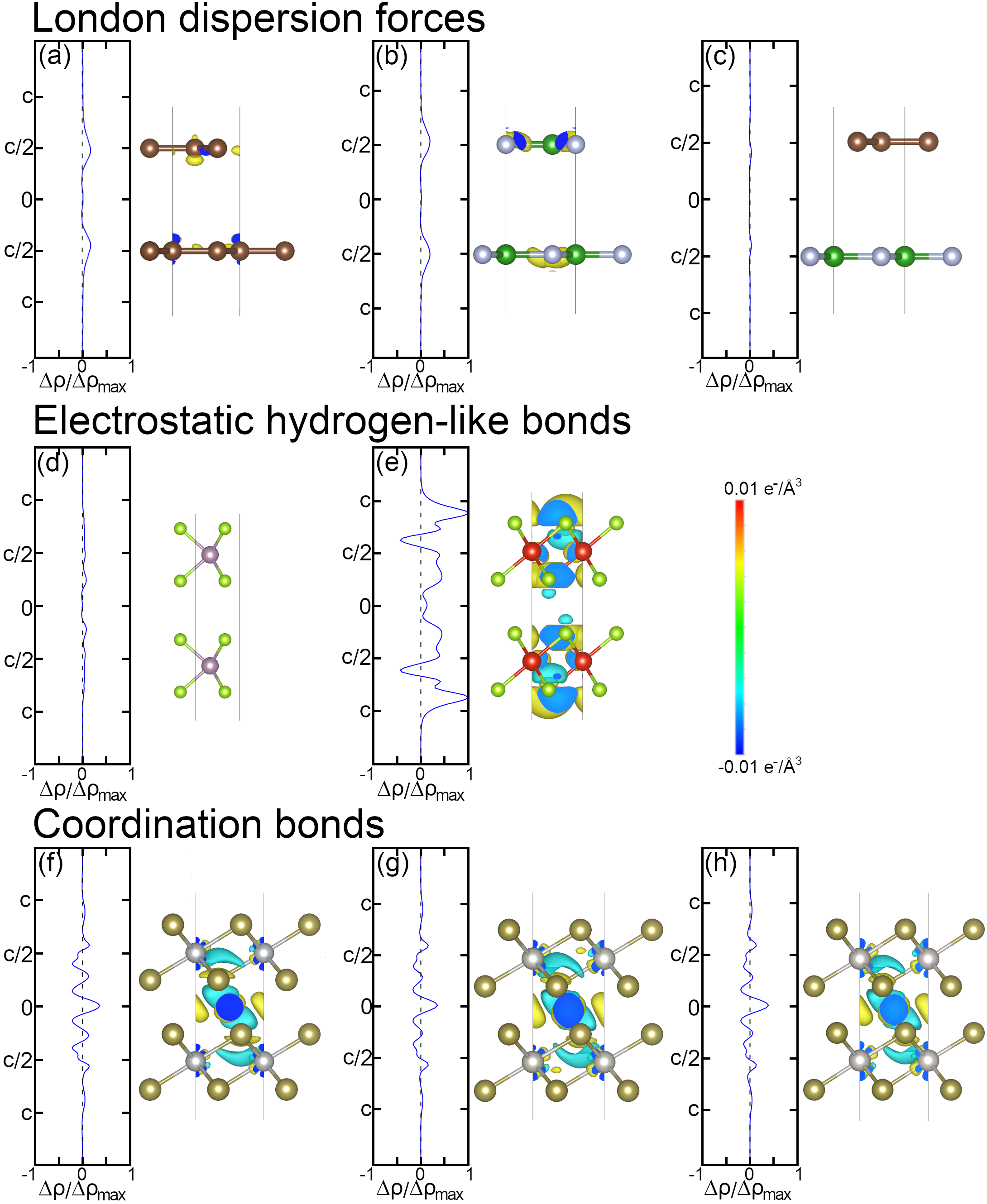}
    \caption{Numerical and visual representations of the  differences between charge densities of two-layer and monolayer phases of graphene (a), hexagonal boron nitride (b), layer of h-BN alternating with layer of graphene(c), H-MoSe$_2$ (d), T-VSe$_2$ (e), NiTe$_2$ (f), PdTe$_2$ (g), PtTe$_2$ (h). Note that the background cut-off, ${\rm BG}=5\times10^{-3} e^-/$\AA$^3$~ for charge densities changes presented numerically in graphs and visualized with Vesta package are the same for all panels of this figure and Fig. 2 but different from those presented in Fig. 1. The maximal value $\Delta \rho_{max}$ equals to 1.6 $\times$ 10$^{-2}$ $e^{-}/$\AA.}
    \label{fig3}
\end{figure*}

The relevance of the performed materials classification by the binding energies and interlayer distance can be justified by calculating charge densities, which, as we will show below, provide a more strict classification. For that purpose, we analyze the patterns of the charge density by comparing the cases of multilayer bulk structures and isolated monolayers. Results of such calculations (Fig. \ref{fig2}) shown in the same scale on the charge density for all the considered systems perfectly coincide with proposed three distinct types to classify the considered systems. Formation of the interlayer bonds in systems we associate with London dispersion forces is accompanied by a tiny redistribution of the charge densities, InSe and $\alpha$-2D-silica in the interlayer space (Fig. \ref{fig1}d-f and Fig. 5). $\Delta \rho$ for graphene and similar systems is more than one order magnitude smaller than in diselenides and ditelurides  (see Fig. \ref{fig2} where we use the same scale for all the investigated systems). The obtained patterns of charge redistribution for graphene, hBN, h-BN are also different from that calculated for traditional van der Waals systems discussed above (see Fig. \ref{fig1}a-c vs \ref{fig1}d-f). Therefore, we can conclude that in addition to London dispersion forces the interactions between occupied and/or unoccupied $\pi$-orbitals of $sp2$-hybridized atoms also take place and these systems can be discussed as van der Waals\cite{Israelachvili2011} only because there is no a special name for this particular type of bonds. The overlap of the orbitals clearly seen in patterns of charge densities between layers of graphite (Fig. \ref{fig1}d) unveils the nature of peculiarities of c-axis conductivity observed in graphites \cite{Fu2015,Zhang2016,Wei2014,TSANG1976}. Note that in the case of bilayer the charge redistribution after formation of interlayer bonds is more visible (see Fig. \ref{fig3}a-c), which can be related with violation of periodicity that could affect both charge fluctuations and $\pi$-$\pi$-interactions. In addition, unusual properties of graphene bilayer discussed since the first years of graphene theoretical and experimental studies \cite{Nilsson2008,Meyer2007} partially related with this uncommon feature of in–plane charge redistribution. 

Electrostatic hydrogen-like bond manifests itself by small green and blue spots above and below the layers shown in Fig. \ref{fig2}d-e and Fig. \ref{fig5}b,d, which can be considered as a mutual polarization of each layer with an effective field of other layers. Breaking periodicity in two-layer structures results in asymmetry of effective electric field acting on each layer that leads to a visible rearrangement of the charge patterns in these systems  (Fig. \ref{fig3}d-e). Importantly, in the case of VSe$_2$-bilayer the charge redistribution is much larger than that in MoSe$_2$ bilayer and bilayers of the studied ditelurides (see Fig. \ref{fig3}d and Fig.\ref{fig3}f-h). This peculiarity of VSe$_2$ can be caused by unique electronic structure of this material\cite{Boukhvalov2020,Pushkarev2019} characterized by single electron in $3d$ shell of VSe$_2$ and could be a reason for observed anomalous structural behaviour of VSe$_2$ bilayers\cite{Esters2017,Chen2020}. In contrast to the weak London dispersion forces bonds and moderate hydrogen-like electrostatic bonds discussed above, formation of coordination-like bonds is associated to the overlap between occupied $d$ orbitals of transitional metals and unoccupied $5p$ orbitals of Te. Presence of such coordination bonds leads to an increase of electron density in the central parts of the interlayer space in ditelurides. Here the main indicator is a distinct blue spot in the center of the interlayer space in Fig. \ref{fig2}f-h , \ref{fig3}f-h and Fig. 5b,d. We would like to point that appearance of the visible clouds of electron density in the interlayer space  observed in dichalcogenides but not in graphene and hBN is one of essential conditions for the formation of charge density waves observed these systems \cite{Yang2014,Feng2020,Eaglesham1986,Barja2016,Xu2021}. 

\begin{figure*}[t]
    \centering
    \includegraphics[width=0.99\columnwidth]{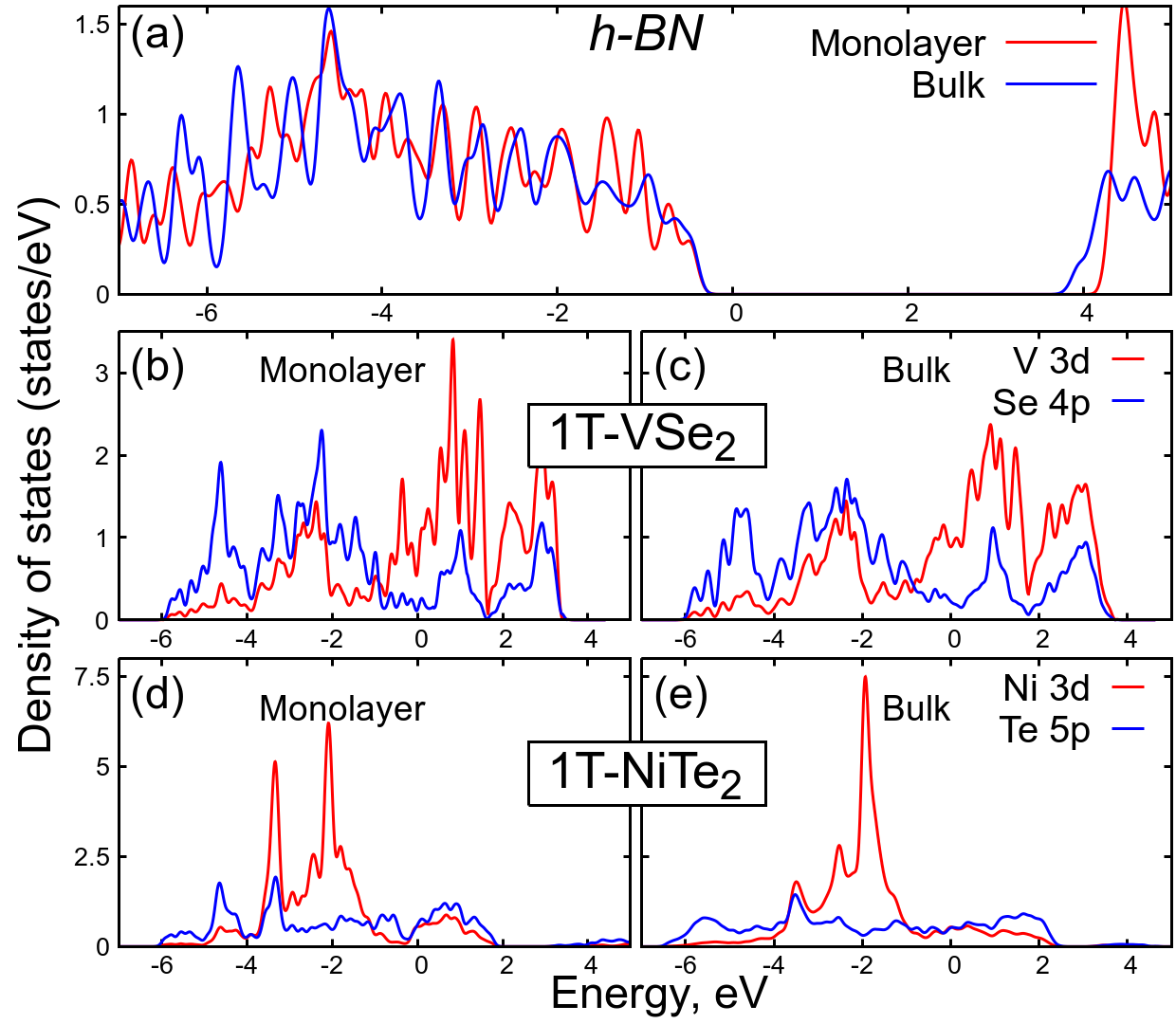}
    \caption{Total densities of states of hBN (a), and partial densities of states calculated for monolayer VSe$_2$ (b) and bulk VSe$_2$ (c) and NiTe$_2$ (d,e).  The Fermi energy is equal to zero.}
    \label{fig4}
\end{figure*}

\begin{figure*}[t]
    \centering
    \includegraphics[width=0.99\columnwidth]{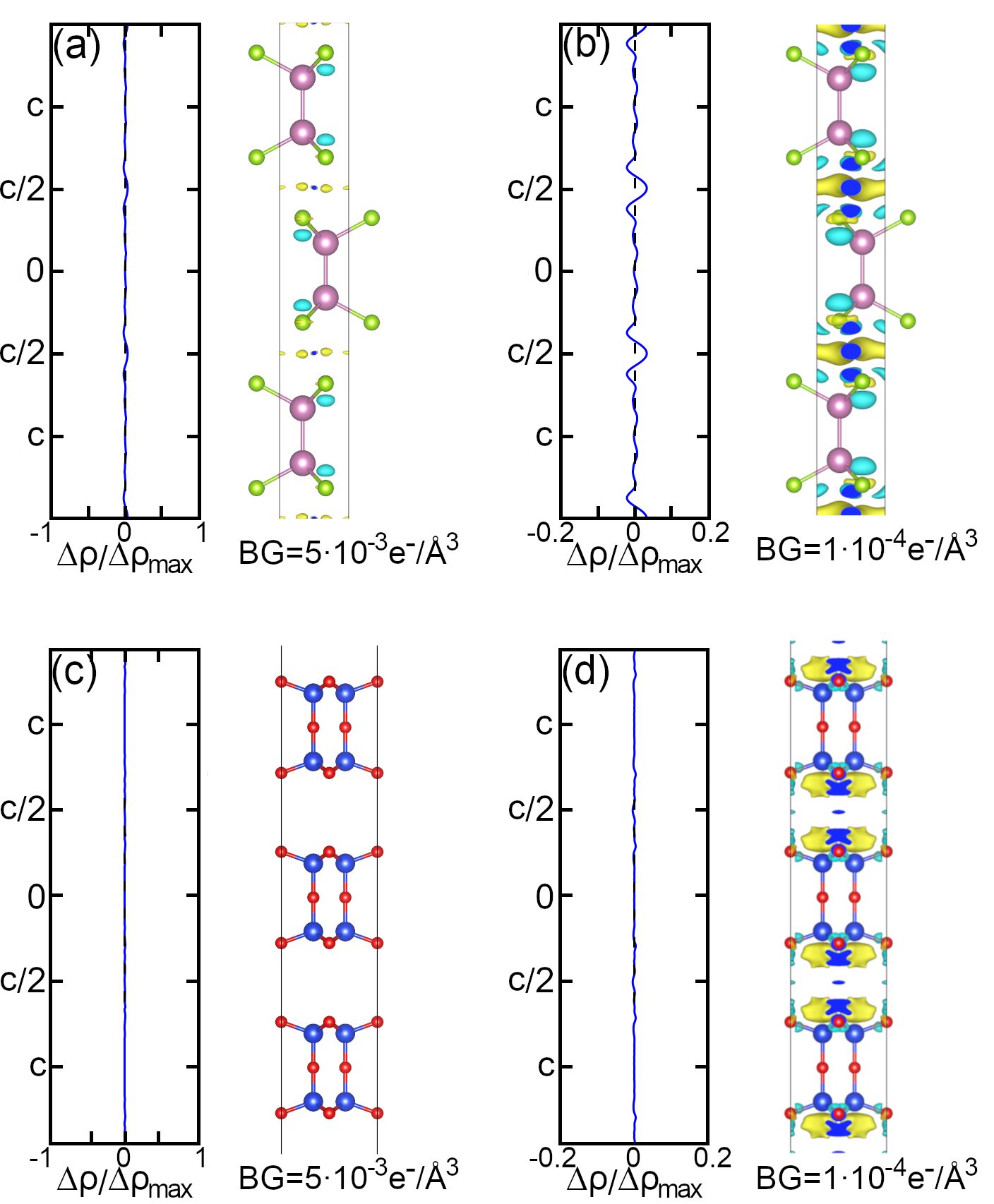}
    \caption{Numerical and visual representations of the differences between charge densities of bulk and monolayer phases of InSe (a,b) and $\alpha$-SiO$_2$ (c,d) with different background cut-off value. The maximal value $\Delta \rho_{max}$ equals to 1.6 $\times$ 10$^{-2}$ $e^{-}/$\AA.}
    \label{fig5}
\end{figure*}

The difference between the considered systems in the particular interlayer bond type can be also traced out on the level of the calculated densities of states. Formation of weak van der Waals bonds does not lead to visible changes in the electronic structure of hBN. In the case of formation electrostatic bonds there is only a tiny shift of the bands caused by appearance of an external electric field from other layers. (Fig. \ref{fig4}b-c). Contrary, formation of the  coordination bonds leads to visible changes in electronic structure of both cations and anions in ditelurides (Fig. \ref{fig4}d-e). 

The differences in the types of interlayer bonds between diselenides and ditelurides can be explained by a significant variation of ionic radii in these systems. For transitional metals coordinated with six neighbours the value of ionic radii vary insignificantly between 0.48 and 0.66 {\AA} even for platinum. In contrast to metal centers the values of ionic radii of anions in dichalcogenides is 1.84, 1.98 and 2.21 {\AA} for S, Se and Te, respectively. Thus, in disulphides and even in diselenides, the size of anion is not enough to overlap with cation from other layers and form coordination bond. To visualize this difference we plot local electrostatic potential for mono- and bilayers of  graphene, MoSe$_2$ and NiTe$_2$ (see Fig. 6a-c). This picture also demonstrates a visible difference between three different systems and especially between diselenides and ditelurides. In the case of graphene and MoSe$_2$ the potential goes to zero at the central point between layers. Contrary in the case of NiTe$_2$-bilayer some nonzero potential observed even in the intermediate region between the layers. The most vivid difference in overall patterns of the potential takes place for systems with dative bonds and other types. In the case of graphene and MoSe$_2$ the potential curves are characterized by distinct kinks on the edges (see Fig. 6a,b). To depict these kinks, we calculate and plot the first derivative of the potentials (see Fig. 6e-f). In these figures one can clearly see kinks at the edges of potentials of graphene and MoSe$_2$ which behave similar to vertical lines on the level of the first derivative. Thus, such a visualization of the potential and calculation of the first derivative for the potentials can be used as alternative approach to characterization of the type of non-covalent bonds in layered systems.

Following the analysis presented above, we can refer graphene, hBN and diselenides as two-dimensional systems with different types of interlayer bonds in contrast to ditelurides, which can be considered as quasi–2D systems with coordination bonds between layers. Hence formation of the surface in ditelurides is related to breaking coordination bonds, which makes the surface of ditelurides to be similar to the surface of bulk of 3D materials. Presence of some analog of dangling bonds on the surface of ditelurides results in chemical instability (rapid oxidation at ambient conditions) of these materials reported in experimental works \cite{Nappini2020,D'Olimpio2020_2} in contrast to chemical stability of diselenides (see, for example, Ref. \citenum{Edla2020}). Another 3D-like property of ditelurides is dependence of the work function on the number of layers. Results of our calculations (see Table \ref{tab2}) demonstrate insignificant difference between work functions of mono- and bilayers for systems with van der Waals interlayer bonds and dichalcogenides. For 
$\alpha$-2D-SiO$_2$ and InSe difference between WF calculated for mono and bilayers is larger, which can be related with more significant influence of electrostatic interlayer interactions on the covalent bonds inside layers (see Fig. 5).  For all the considered diteluride systems the difference between $\Psi_{\rm work}$ calculated for mono- and bilayer is about 0.3 eV. This change of WF with the change of number of layers is similar to that previously observed for slabs of transitional metals for which formation of the surface also is accompanied by the break of dative and even covalent bonds.\cite{Singh-Miller2009} It is important that the shape of local potential curves of NiTe$_2$ is also different from the MoSe$_2$ and graphene (see Fig. 6). In the case of NiTe$_2$ absence of the kinks, which is confirmed by fact that there are no vertical lines in the plot of the first derivative of the local potential, can be explained by forming dative bonds inside bilayer and breaking these bonds on the surfaces of mono- and bilayer.

\subsection{Role of the van der Waals corrections}
At the last step of our investigation we are going to examine the importance of inclusion of the van der Waals interactions (D3 corrections) \cite{VdW,Tkatchenko2009,vonLilienfeld2004} when calculating   atomic structure of layered materials with hydrogen-like or dative interlayer bonds. Following to the work by Tkatchenko et al. \cite{Tkatchenko2009}, such corrections to standard GGA must be considered to define correct values of intermolecular  distances and binding energies even for the interactions between two water molecules connected with hydrogen bonds.  Additionally, in our case the values of the interlayer distances calculated with taking D3 corrections into account are in perfect agreement with experimental values (see Table \ref{tab1}). Thus, the effect on inclusion of the D3 correction should be assessed. To this aim, we performed optimization of the atomic structures of all the considered systems without D3 corrections and compare values of the binding energies and interlayer distances with the values calculated with taking into account D3 corrections. Results of the calculations presented in Table \ref{tab3} demonstrate that for 2D systems in which layers bonded by London dispersion forces (graphene, hBN) the calculations without D3 corrections revealed an overestimation of interlayer distance by about 0.5 \AA. In the case of materials with electrostatic hydrogen-like interlayer bonds (diselenides, InSe) the effect on interlayer distances is more pronounced. In the case of materials characterized by the dative interlayer bonds (ditelurides), calculations without D3 corrections show the increase of interlayer distance by about 0.3 \AA~or less. A similar trend is also observed for the difference in the value of binding energies calculated with and without D3 corrections (see Table \ref{tab3}). It is important that the values of binding energies calculated for the systems with van der Waals of electrostatic interlayer bonds are rather small (3.3 meV/bond or even smaller). Contrary in the case of the systems with dative interlayer bonds the values calculated without D3 corrections remain one order larger than those calculated for other considered systems. Based on these results we can conclude that the basic approximations realized in standard DFT-GGA scheme leads to underestimation of contributions not only from London dispersion forces but also hydrogen-like and even dative bonds and hence taking into account D3 corrections is essential for simulation of the systems we consider.

\section{Conclusions}
In summary, having compared interlayer distances, binding energies, charge densities and electronic structures calculated for layered structures and individual monolayers, we conclude that only the layers in graphite, few–layer graphene and hexagonal boron nitrite are bonded by London dispersion forces and may well be referred as van der Waals systems. In contrast to these flat monolayers, the diselenides, VSe$_2$ and MoSe$_2$ demonstrate a significant redistribution of the charge density upon formation of bulk or bilayer. Benchmarking the calculated properties of diselenides with data known from the literature reveals an essential role of electrostatic interaction between neighbouring layers. Transition from diselenides to ditelurides characterized by similar atomic structures is accompanied an increase of the ionic radii of cations, which makes possible additional coordination of metal centers by orbitals of cations belonging to other layer(s). As the result so-called dative or coordination chemical bonds between neighbouring layers are formed in ditelurides. Such a switch from electrostatic to coordination types of the interlayer bonds leads to visible changes in electronic structure and work function of ditelurides within monolayer/bulk transition in contrast to negligible changes of these properties in diselenides. In ditelurides coordination interlayer bonds are responsible for amplifying three-dimensional character of inter-atomic couplings in these systems. The same coordination-like inter-layer interactions also contribute to the formation of unsaturated chemical bonds on the surface, which explains experimentally observed chemical instability of ditelurides. A redistribution of the charge density upon formation of electrostatic or coordination interlayer bonds can be the key to understanding of charge density waves observed in multiple dichalcogenides. 
The proposed scheme for recognizing the leading type of interlayer bonds is computationally cheap and can be used for diagnosing a large number of real and hypothetical materials with further prediction of their chemical and physical properties.  

\section*{Conflicts of interest}
There are no conflicts to declare.

\section*{Acknowledgement}
This work was supported by the Russian Science Foundation, Grant No. 21-72-10136.

%%%END OF MAIN TEXT%%%

\newpage

\begin{table}[ht]
\begin{tabular}{c| c c c c c c}
\hline
 &$E_{\rm bind}$, &$a_{\rm 1L}$ \AA &$a_{\rm 2L}$, \AA &$a_{\rm B}$, \AA &$d_{\rm 2L}$, \AA &$d_{\rm B}$, \AA \\
  & meV/bond & & & & &  \\
\hline

graphite(AB)         & 17.2 & 2.47  & 2.47 & 2.47 & 3.40& 3.41 \\
& &$e$- 2.46\cite{t19} & $e$- 2.46\cite{t6} & $e,t$- 2.46\cite{t2,t5} & $e$- 3.35\cite{t6} & $e,t$- 3.50\cite{t1,t2} \\

hBN(AB)          & 16.3  & 2.51  & 2.51 & 2.51 & 3.34 & 3.36 \\
& & $e$- 2.49\cite{t7} &  &$e$- 2.50\cite{t8} & &$e$- 3.33\cite{t8} \\

hBN/graphene(AB)   & 16.4  & ---   & 2.49   & 2.49  &  3.48  & 3.52 \\
& & & & $t$- 2.46 \cite{t13} & & $t$- 3.49\cite{t14} \\

$\alpha$-SiO$_2$          & 30.2  & 5.31  & 5.30 & 5.26 & 3.37 & 3.19 \\
& & $t$- 5.31\cite{Gao2017,Altman2013} &  & & &  \\

 \emph{2H}-MoSe$_2$(AB)& 47.2 & 3.29 & 3.29& 3.29 & 3.18 &  3.17 \\
& & & &$e$- 3.28\cite{Agarwal1986} & &$e$- 3.22\cite{Agarwal1986}\\

InSe(AB)          & 41.3  & 4.06  & 4.06 & 4.07 & 3.20 & 3.10 \\
& & $t$- 4.08\cite{Zolyomi2014,Wu2018} &  &$e$- 4.05\cite{Man1976} &$t$-3.18\cite{Wu2018}&  \\

 \emph{1T}-VSe$_2$(AA)& 48.3  & 3.32 & 3.30 & 3.32 & 3.05 & 3.14  \\
& & $e$- 3.31\cite{t10,t11} & & $e$- 3.33\cite{t11} & &$e$- 3.14\cite{t11} \\

$1T$-NiTe$_2$  & 119.5  & 3.77 & 3.82  & 3.88 & 2.64 & 2.47 \\
& & $t$- 3.77\cite{t15} & & $t$- 3.90\cite{t16}  &  & $t$- 2.65\cite{t16} \\

 \emph{1T}-PdTe$_2$  & 136.3 & 3.99 & 4.03 & 4.07 & 2.34 & 2.27 \\
& & $e$- 4.04\cite{t12} & &$e$- 4.03\cite{Kim1990} & &  \\ %$$e$- 2.57\cite{t12}

 \emph{1T}-PtTe$_2$  & 112.5 & 3.98 & 4.02 & 4.07 & 2.39 & 2.34\\
& & $t$- 4.07\cite{t17} & & $e,t$- 4.03\cite{Kjekshus1959,t17}  & &  \\ %$e$- 2.56\cite{Kjekshus1959}
\hline
\hline
\end{tabular}
\caption{Calculated binding energy per bond (see section 2 in the text) and lattice parameters in different configurations of considered structures in bulk (B), bilayer (2L) and monolayer (1L) forms. $d_{\rm B}$ and $d_{\rm 2L}$ denote the inter-layer distances in bulk and two-layer structures, respectively.  The values denoted with $e$ and $t$ labels correspond to those obtained in previous experiment and theoretical works.}
\label{tab1}
\end{table}

\begin{table*}[ht]
\begin{tabular}{c| c c }
\hline
 &$\Psi_{\rm work}$ 1L, eV &$\Psi_{\rm work}$ 2L, eV \\
\hline

graphite(AB)           & 4.53 & 4.29  \\

hBN                    & 5.78 & 5.74\\

hBN/graphene           & 4.72 & 4.72\\

$\alpha$-SiO$_2$ & 7.85 & 7.79\\

\emph{2H}-MoSe$_2$ (AB)& 5.30 & 5.30\\

InSe(AB) & 5.67 & 5.39\\

\emph{1T}-VSe$_2$ (AA) & 5.00 & 5.03\\

$1T$-NiTe$_2$          & 4.39 & 4.54 \\

\emph{1T}-PdTe$_2$     & 4.35 & 4.62 \\

\emph{1T}-PtTe$_2$     & 4.02 & 4.26 \\

\hline
\hline
\end{tabular}
\caption{Calculated work function $\Psi_{\rm work}$ for one- and two-layer structures of the considered systems. The experimental estimate of the work function for graphene is 4.51 eV\cite{t18}.}
\label{tab2}
\end{table*}

\begin{table*}[ht]
\begin{tabular}{c| c c c c }
\hline

 &E$_{\rm bind}$ with &E$_{\rm bind}$ without&d$_{\rm interlayer}$ &d$_{\rm interlayer}$ \\
  &  vdW, meV/bond &  vdW, meV/bond & with vdW, \AA & without vdW, \AA \\
\hline

graphite(AB)           & 17.2 &0.2 &3.41 &3.93 \\

hBN                    & 16.3 &0.2 &3.36 &3.82\\

hBN/graphene           & 16.4 & 3.3 &3.52 &4.06\\

$\alpha$-SiO$_2$ &30.2& 0.9 &3.19 &3.34\\

\emph{2H}-MoSe$_2$ (AB)& 47.2 & 0.9 &3.18 &4.00\\

InSe(AB) & 41.3 & 1.3 &3.10 &3.85\\

\emph{1T}-VSe$_2$ (AA) & 48.3  & 0.8 &3.14 &3.69\\

$1T$-NiTe$_2$          & 119.5 & 41.0 &2.47& 2.77\\

\emph{1T}-PdTe$_2$     & 136.3 & 47.7 &2.27& 2.47\\

\emph{1T}-PtTe$_2$     & 112.5 & 17.4 &2.34 &2.63\\

\hline
\hline

\end{tabular}
\caption{Calculated binding energy E$_{\rm bind}$ and interlayer distances d$_{\rm interlayer}$ for bulk materials with and without adding van der Waals forces into account}
\label{tab3}
\end{table*}

\begin{figure*}[t]
    \centering
    \includegraphics[width=0.99\columnwidth]{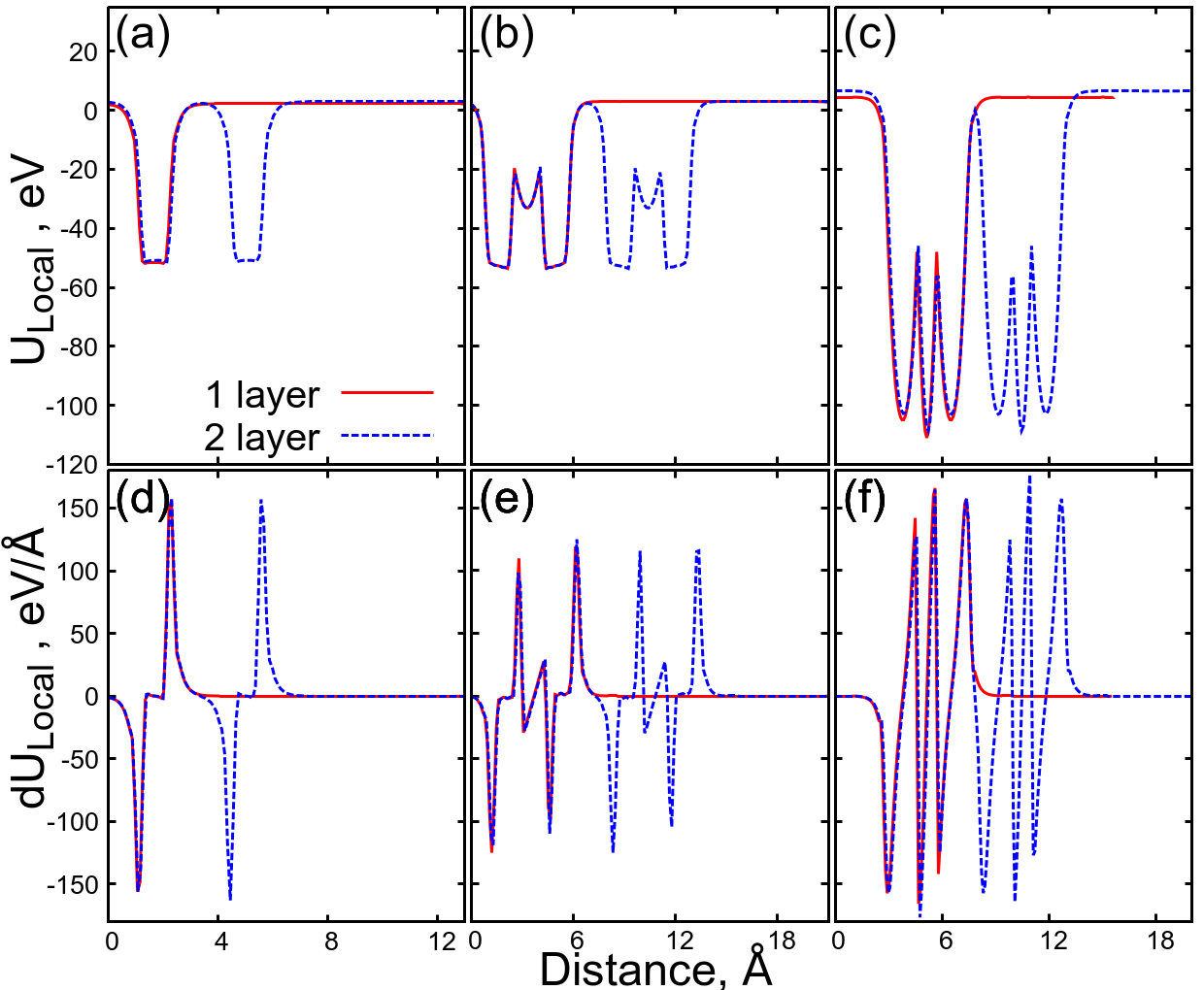}
    \caption{Local potential U$_{\text{\normalfont Local}}$ functions  via distance and first derivative of local potential dU$_{\text{\normalfont Local}}$/dL  for monolayer (red line) and twolayer (blue line)  graphene (a,d), \emph{2H}-MoSe$_2$ (b,e), $1T$-NiTe$_2$ (c,f) respectively.}
    \label{fig6}
\end{figure*}

\newpage 

%%%REFERENCES%%%
\bibliography{main}

\end{document}